\documentclass{iopart}
\usepackage{graphicx}% Include figure files

\begin{document}
%------------------------------------------------------------------------
%                            Title and Abstract
%------------------------------------------------------------------------
\title{Theory of large-scale matrix computation \\ 
and applications to electronic structure calculation}
\author{T. Fujiwara$^{1,2}$, T. Hoshi$^{3,2}$ and S. Yamamoto$^2$}
\address{(1) Center for Research and Development of Higher Education, 
The University of Tokyo, 
Bunkyo-ku, Tokyo, 113-8656, Japan}
\address{(2) Core Research for Evolutional Science and Technology, 
Japan Science and Technology Agency (CREST-JST), Japan}
\address{(3) Department of Applied Mathematics and Physics, 
Tottori University, Tottori 680-8550, Japan}
%------------------------------------------------------------------------
%                              abstract
%------------------------------------------------------------------------
\begin{abstract}
We review our recently developed methods 
for large-scale electronic structure calculations, 
both in one-electron theory and many-electron theory. 
The method are based on the density matrix representation, together with 
the Wannier state representation and the Krylov subspace method, 
in one-electron theory of a-few-tens nm scale systems. 
The hybrid method of quantum mechanical molecular dynamical simulation 
is explained.
The Krylov subspace method, the CG (conjugate gradient) method and 
the shifted-COCG (conjugate orthogonal conjugate gradient) method, 
can be applied to the investigation of the ground state and the excitation 
spectra in many-electron theory. 
The mathematical foundation of the Krylov subspace method for large-scale 
matrix computation is focused and   
the key technique of the shifted-COCG method, e.g. 
the collinear residual and seed switching, is  explained.  
A wide variety of applications of these extended novel algorithm  
is also explained. 
These are the fracture formation and propagation, liquid carbon and 
formation process of gold nanowires, 
together with the application to the extend Hubbard model. 
\end{abstract}

\date{\today}
\maketitle
%------------------------------------------------------------------------
%                              Introduction
%------------------------------------------------------------------------
\section{Introduction}

Large-scale matrix computation is crucial in the electronic structure theory, 
both in one-electron theory for large-scale systems and in many-body theory 
for strongly interacting electron systems. 
Interplay of the electronic structure and nano-scale atomic structure 
plays an essential role in physical properties of nanostructure materials 
and the Order-N algorithm has been extensively investigated. 
The size of the Hilbert space grows exponentially with linear increase of the system size 
in many-electron problems. 
Very important ten algorithms were invented in the 20th century.~\cite{CSE,SIAM}
These algorithms are the Krylov subspace method, 
the QR algorithm, the Householder algorithm, 
the Fast Fourier Transformation (FFT) etc, 
where the most of them are of the matrix algebra and the order-N algorithm. 
The FFT algorithm is one of the basis of the local density approximation (LDA) 
in the density functional theory (DFT) 
and the Lanczos method, one of the Krylov subspace method, is that of 
the many-body electron theory. 
The efficiency of the modern Krylov subspace method seems not to be widely known 
in the field of electron theory, both in LDA and many-electron theory. 
In one-electron theory or DFT, the primarily important states are 
the states near the Fermi energy or the band gap.  
Then the standard mathematical tool is the diagonalization of 
the Hamiltonian matrix. 
This may be a serious difficulty in large scale systems. 
In many electron theory, the difficulty is the huge size of the 
Hamiltonian matrix and the resultant memory size and computational time. 
These are just the targets of the field of the large-scale matrix computation 
mentioned above. 
In this paper, we report our recent activity in 
(1) developing the quantum mechanical molecular dynamical (MD) simulation method 
with the exact diagonalization, the Wannier states representation and 
the Krylov subspace method in nano-scale systems up to a few $10$~nm size 
and 
(2) the investigation of many-electron problem, {\it i.e.} 
the degenerated orbital extended Hubbard Hamiltonian 
of the size of $6.4\times 10^7$, with the Krylov subspace method. 
We explain the key aspects in one-electron theory in a large-scale system 
and the many-electron theory in Sect. II. 
Section III is devoted to the explanation of the Krylov subspace methods. 
Several applications are reviewed in Sect. IV and 
the summary is given in the last section. 
%%  

%====================================================================
\section{One-electron theory {\it vs.} many-electron theory}

%--------------------------------------------------------------------
\subsection{One-electron spectrum in large-scale systems}

%--------------------------------------------------------------------
\subsubsection{Density matrix formulation} \ \

The LDA calculation is  based on the variational principles 
and usually on eigen-function representation of the ground state. 
However, the eigen-functions are not always necessary in actual calculation 
nor useful in numerical investigation 
of large-scale systems. 
Instead, one can construct the formulation with the one-body density matrix.~\cite{KOHN96}
Any physical property can be represented by the density matrix $\rho$ as
\begin{eqnarray}
\langle {\hat X} \rangle = {\rm Tr} [\rho {\hat X}] = \sum_{ij} \rho_{ij} {\hat X}_{ij},
\end{eqnarray} 
where ${\hat X}$  is an operator of the physical property $X$ and 
$i$ and $j$ denote atomic sites and orbitals. 
Energy and forces acting on an individual atom can be calculated by replacing ${\hat X}$ 
by the Hamiltonian or its derivative. 
Therefore, 
one needs only $(i,j)$ elements of the density matrix $\rho$ 
corresponding to non-zero $X_{ij}$  
but not all elements. 
The density matrix $\rho_{ij}$ is given as 
\begin{eqnarray}
\rho = \sum_{\alpha}^{\rm (occ)} |\phi_\alpha \rangle \langle\phi_\alpha|   , 
\label{eq:densityM-1}
\end{eqnarray}
where $|\phi_\alpha \rangle$ is the eigenstates or the Wannier states and 
the summation is restricted within the occupied states. 
It can be also written as 
\begin{eqnarray}
 \rho_{ij} = -\frac{1}{\pi}\int_{-\infty}^{+\infty} {\rm d}\varepsilon 
 {\rm Im}G_{ij}(\varepsilon) f\Big( \frac{\varepsilon - \mu}{k_{\rm B}T} \Big) ,
\label{eq:densityM}
\end{eqnarray}
where $G_{ij}$ is the Green's function defined as 
\begin{eqnarray} 
G_{ij}(\varepsilon) = [(\varepsilon + {\rm i}\delta -{\hat H})^{-1}]_{ij} .  
\label{eq:gf}
\end{eqnarray}
Here,  $\mu$, $k_{\rm B}$, $T$ and $f$ are the chemical potential, 
the Boltzmann constant, the temperature and 
the Fermi-Dirac distribution function, respectively. 
We have developed a set of computational methods 
for electronic structure calculations, {\it i.e.}
the generalized Wannier-state method,~\cite{HOSHI2000, HOSHI2003,HOSHI2005}
the Krylov subspace method (the subspace diagonalization method~\cite{TAKAYAMA2004} 
and the shifted COCG method~\cite{TAKAYAMA2006}) 
and the generalized Wannier-state solver with parallelism.~\cite{geshi2004} 
These methods are ones for calculating 
the one-body density matrix and/or the Green's function 
for a given Hamiltonian.
Calculation was carried out
using the tight-binding formalism of the Hamiltonian. 
These methods can be used in a  hybrid way 
as is explained in \ref{multisolver}.~\cite{HOSHI2006}   
%% 

%--------------------------------------------------------------------
\subsubsection{Wannier state representation}\ \ 

The Order-N algorithm can be constructed in semiconductors and insulators 
on the basis of the Wannier state representation. 
The generalized Wannier states 
are localized wavefunctions in condensed matters 
obtained by the unitary transformation 
of occupied eigenstates,~\cite{MAURI, MARZARI, HOSHI2000}
and also obtained by an iterative way, 
starting a trial localize wavefunctions, with a mapped eigen-value equation~\cite{HOSHI2000} 
\begin{eqnarray}
 H_{\rm WS}^{(i)}  | \phi _i^{\rm (WS)} \rangle 
 = \varepsilon_{\rm WS}^{(i)}  | \phi _i^{\rm (WS)} \rangle,
 \label{MFE}
\end{eqnarray}
where 
\begin{eqnarray}
& &  H_{\rm WS}^{\rm (i)} \equiv 
 H  + 2 \eta_{\rm s} \bar{\rho}_i - H  \bar{\rho}_i - \bar{\rho}_i H  
 	 \label{EQ-WANI-HAMI} \\
& & \bar{\rho}_i 
     \equiv  \rho -  | \phi_i^{\rm (WS)} \rangle \langle \phi_i^{\rm (WS)} | 
	 =\sum_{j (\ne i)}^{\rm occ.} | \phi_j^{\rm (WS)} \rangle \langle \phi_j^{\rm (WS)} |,  
\end{eqnarray}
and the energy parameter $\eta_{\rm s}$ should be much larger
than the highest occupied level. 
Once one obtains the Wannier states, the density matrix can be easily constructed 
by Eq.~(\ref{eq:densityM-1})
and the force acting on each atom can be calculated. 
We observed that the bond forming and breaking processes are 
well described in the localized Wannier states as changes
between a bonding and non-bonding orbital. 
The Wannier states depend upon the local environment and 
the above iterative procedure is suitable to the MD simulation. 
%% 

%--------------------------------------------------------------------
\subsubsection{Krylov subspace method}\ \ 

In metallic systems, the Krylov subspace method is very useful 
to achieve the computational efficiency (accuracy and speed).~\cite{TAKAYAMA2004,TAKAYAMA2006}
The Green's function can be calculated in the Krylov subspace  
and one calculates the density matrix by Eq.~(\ref{eq:densityM}). 
Details are explained in Sec.~\ref{sec:Krylov}. 
The Krylov subspace method is, of course, applicable 
to semiconductors and insulators, too.
%%

%--------------------------------------------------------------------
\subsubsection{Comparison among solver methods and Order-N character}\ \ 

\begin{figure}[htbp] 
\begin{center}
\resizebox{0.48\textwidth}{!}{
  \includegraphics{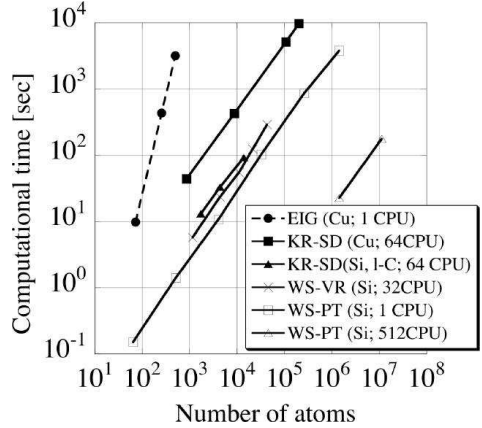}
}
\end{center}
\caption{\label{FIG-BENCH} 
The computational time
as a function of the number of atoms ($N$). 
\cite{HOSHI2003, HOSHI2005, HOSHI2006}
The time was measured 
for metallic (fcc Cu and liquid C) and insulating (bulk Si) 
systems with up to 11,315,021 atoms, 
by the conventional eigenstate calculation (EIG) 
and by our methods for large systems; 
KR-SD, WS-VR and WS-PT methods.  
See the original papers~\cite{HOSHI2003, HOSHI2005, HOSHI2006}
for the details of parallel computation.
}
\end{figure}

Figure \ref{FIG-BENCH}  demonstrates 
our methods for 10$^2$-10$^7$ atoms 
with and without parallel computation,~\cite{HOSHI2003,HOSHI2005,HOSHI2006}
where the computational time is shown 
for the standard eigen-state solver (EIG) and our developed solver methods, 
Wannier-state solver with variational procedure (WS-VR), 
Wannier-state solver with perturbation procedure (WS-PT) 
and Krylov subspace solver with subspace diagonalization (KR-SD).
Parallel computations are achieved by 
the Open MP technique (http://www.openmp.org). 
The Hamiltonian forms used here are 
the Slater-Koster-form ones of silicon~\cite{KWON1994} and carbon~\cite{XU},  
the linear-muffin-tin orbital (LMTO) theory~\cite{LMTO} 
in a form of the first-order ($H^{(1)}$) for copper. 
Among the data in Fig.~\ref{FIG-BENCH} except
one by the eigen-state solver, 
the computational cost is
\lq order-$N$' or linearly proportional to the system size ($N$), 
up to ten-million atoms 
and shows a satisfactory performance in parallel computation. 
The computational performance of the Wannier-state methods 
can be faster, at best by several hundred times, 
than that of the Krylov subspace method (See Fig.~\ref{FIG-BENCH}, for example), 
particularly, if a dominant part of wavefunctions  are well localized.
Now the program package (`ELSES' = Extra Large-Scale Electronic Structure calculation) 
is being prepared.~\cite{ELSES} 
%%

%%%%%%%%%%% figure 1 %%%%%%%%%%%%%%%%%%%%%%%
%%%%%%%%%%%%%%%%%%%%%%%%%%%%%%%%%%%%%%%

%--------------------------------------------------------------------
\subsubsection{Multiple solver method}\label{multisolver} \ \ 

Since our method based on the density matrix formulation, 
we can construct another very important method ``the multiple solver method''.
The basic idea is the division of the Hilbert space;
\begin{eqnarray}
\rho =\rho_{\rm A}+\rho_{\rm b} , \ \ \  \rho_{\rm A}\rho_{\rm B}=0 ,
\label{multi}
\end{eqnarray}
and the calculation can be done independently on different parts A and B.
The importance is the fact that this hybrid method is completely within 
the quantum mechanical framework. 
Then we can use this hybrid scheme as the multiple solver method 
in nano-scale systems 
and the calculated results does not cause any artificial discontinuity 
of physical quantities which might always occur in the hybrid scheme of 
naive simple division of the physical space. 
The choice or hybrid of the solvers is important 
for an optimal calculation with a proper balance 
between accuracy and computational cost.~\cite{HOSHI2003,HOSHI2006}
%%

%--------------------------------------------------------------------
\subsection{Ground state of many-electron theory and excitation spectra}

In many-electron theory, we usually treat large matrices and 
the calculation becomes more and more difficulty, 
since the occupation freedom of one site grows exponentially 
and the matrix size is extraordinarily large.~\cite{yamamoto2007a}
In order to get the precise eigen-energy and eigen-vector of the ground state, 
one should use the Lanczos method and the CG method (the inverse iteration method) 
simultaneously. 
The Lanczos method is useful to get the approximate eigen-energy and eigen vectors  
of the ground state. 
However, the orthogonality of the generated basis vectors is broken 
at low iteration steps 
and the precision of the ground state energy and wavefunction 
could not be preserved. 
Then we use the CG method (the inverse iteration method) 
to improve them. 
After estimating the accuracy of the calculation (the norm of residual vector), 
we should repeat this procedure till the enough accuracy is obtained.  
After  we obtain the wavefunction of the ground state,  
we should analyze the properties of excitations in a wide range of energy. 
For this purpose, the shift property of the COCG method, 
{\it i,e,} the sifted COCG method, 
can be a powerful tool. 
The advantages of the shifted COCG method is 
an efficient algorithm of solving shifted linear equations,  
the error-monitoring ability 
during the iterative calculation and the robustness. 
Then, it is very suitable for the problems 
in the many-electron problem.~\cite{yamamoto2007b}
The detailed explanation is given in Sec.\ref{sec:Krylov}. 
%%

%------------------------------------------------------------------------
%                              A novel theory for large scale
%------------------------------------------------------------------------
\section{Krylov subspace method}\label{sec:Krylov}
 
%%
%------------------------------------------------------------------------
\subsection{Krylov subspace}\label{subsec:def_Krylov}

We consider the simultaneous linear equations 
\begin{eqnarray}
     && [(\varepsilon + {\rm i}\delta){\bf 1} -{\hat H}]~|x_j\rangle = |j \rangle ,
\label{eq:lin_eq}
\end{eqnarray}
for a given  vector $|j\rangle$, real numbers $\varepsilon$ and $\delta$. 
${\bf 1}$ is the unit matrix. 
When ${\hat H}$ is a huge $N \times N$ matrix, the inverse of ${\hat H}$ 
or $[(\varepsilon + {\rm i}\delta){\bf 1} -{\hat H}]$  is not easily obtained  
or impossible to obtain and the iterative method becomes a useful concept. 
One can  obtain an approximate eigen vector  $|x_j\rangle$ 
in a subspace spanned by 
vectors $\{{\hat H}^n|j \rangle\}$;  
\begin{eqnarray}
 K_\nu \big({\hat H},|j \rangle \big)  
\equiv {\rm span}\big\{ |j \rangle ,\ {\hat H}|j \rangle ,\ {\hat H}^2|j \rangle,\ 
 \cdots \ , \ {\hat H}^{\nu-1}|j \rangle \big\}. 
\label{def:Krylov}
\end{eqnarray}
This subspace $ K_\nu \big({\hat H},|j \rangle \big)$ is called the Krylov subspace.  
The basic theorem of the Krylov subspace is the invariance of the subspace 
under a scalar shift $\sigma {\bf 1}$; 
\begin{eqnarray}
   K_\nu \big({\hat H},|j \rangle \big)
 = K_\nu \big(\sigma{\bf 1}+{\hat H},|j \rangle \big) .
\label{inv_KS}
\end{eqnarray} 
Lanczos found a new powerful way to generate an orthogonal basis for such subspace 
when the matrix is symmetric.~\cite{Lanczos} 
Hestenes and Stiefel proposed an elegant method, 
known as the conjugate gradient (CG) method, 
for systems that are both symmetric and positive definite.~\cite{CGmethod} 
%%

%------------------------------------------------------------------------
\subsection{Subspace-diagonalization}

The first method is to find eigen-vectors $\{ |w_\alpha^{(j)} \rangle \}$ approximated 
in $ K_\nu \big({\hat H},|j \rangle \big)$ 
by diagonalizing the reduced Hamiltonian matrix
\begin{eqnarray}
 H^{K_\nu \big({\hat H},|j \rangle \big)}=\{ \langle K_n^{(j)} | {\hat H} | K_m^{(j)} \rangle \} ,  
\label{eq:Ham_in_KS}
\end{eqnarray}
where $\{| K_m^{(j)} \rangle | m=1,\cdots,\nu \}$ is the orthogonalized basis set of the 
Krylov subspace $ K_\nu \big({\hat H},|j \rangle \big)$, 
which satisfies the three-term recurrence relation and 
constructed by the Lanczos process or the Gram-Schmidt process. 
In this subspace we can calculate the density matrix very easily.~\cite{TAKAYAMA2004} 
The subspace diagonalization method may be accurate enough for the 
several purposes in one-electron spectra in large-scale systems 
and calculation of total and local density of states.  
However, the orthogonality would be broken, when we use a larger number 
of the subspace dimension, 
for the basis vectors satisfying the three-term recurrence relation.~\cite{TAKAYAMA2004,TAKAYAMA2006}
Therefore, the numerical accuracy may be limited 
when one need finer structure of spectra and 
we should extend the methodology to the shifted COCG method. 
In many-electron theory, the Lanczos method is widely used for 
obtaining the eigen-energy and many-electron wavefunction of the ground state in 
the exact diagonalization method. 
The accuracy can be greatly improved 
when we use the CG method. 
%%

%------------------------------------------------------------------------
\subsection{Shifted-COCG method and seed-switching technique}

When the matrix $(\varepsilon_0{\bf 1} -{\hat H})$ is real symmetric, 
then one can use the CG method for an iterative solution 
of the simultaneous linear equation 
$(\varepsilon_0 {\bf 1} -{\hat H}){\bf x}={\bf b}$. 
One should introduce the infinitesimal small (but finite) 
imaginary number ${\rm i}\delta$ 
for the Green's function and the matrix 
$(\varepsilon_0 + {\rm i}\delta){\bf 1} -{\hat H}$ is complex symmetric.  
Then we can use the conjugate orthogonal conjugate gradient (COCG) method 
for solving the equation
$\{(\varepsilon_0+ {\rm i}\delta) {\bf 1} -{\hat H}\}{\bf x}={\bf b}$.~\cite{TAKAYAMA2006} 
Since the energy parameters $\varepsilon$ are arbitrary given 
or continuously changing in a wide energy range, 
one should solve also the shifted linear equations 
\begin{eqnarray}
     [(\varepsilon_0 +\sigma + {\rm i}\delta){\bf 1} -{\hat H}]~|x_j^{(\sigma)}\rangle = |j \rangle ,
\label{eq:shift_lin_eq}
\end{eqnarray}
with a fixed energy (seed) $\varepsilon_0$.
The energy shift parameter $\sigma$ can be even complex. 
The shifted-COCG method was constructed,~\cite{TAKAYAMA2006,yamamoto2007b}
in which the theorem of collinear residual~\cite{FROMMER2003} for the shifted linear systems 
is applied to the COCG method. 
The essential property is based on the basic invariance theorem of the Krylov subspace 
Eq.~(\ref{inv_KS})
under an energy shift $\varepsilon_0 + \sigma$ from  $\varepsilon_0$. 
Therefore, the Krylov subspace for the equation 
$[(\varepsilon_0 +\sigma + {\rm i}\delta){\bf 1} -{\hat H}]~|x_j^{(\sigma)}\rangle = |j \rangle$ 
can be generated from that of 
$[(\varepsilon_0 + {\rm i}\delta){\bf 1} -{\hat H}]~|x_j\rangle = |j \rangle$ 
of a selected seed energy  $\varepsilon_0$. 
The very important fact is that this shift procedure is scalar linear calculation.   
Essential cost for solving Eq.~(\ref{eq:lin_eq}) should be paid only 
for the seed energy $\varepsilon_0$ and the rest is a scalar linear calculation which 
is negligible.  
The choice of the seed energy is not unique and sometime 
the calculations cannot be finished under a required criteria. 
Then one should continuously change the energy parameter 
and choose a new seed energy $\varepsilon + i\eta$ again. 
Essentially important point is that 
we can continue the calculation with a new seed energy, 
keeping calculated information of the former seed energy. 
This is another very important property called 
\lq seed-switching'.~\cite{SOGABE2007,yamamoto2007b} 
%%

%------------------------------------------------------------------------
\subsection{Accuracy control with residual vector and  robustness of 
shifted COCG method}

It is essentially important to know the accuracy of the solution 
during the iteration procedure and 
we can monitor the convergence behavior of the iterative solutions 
of the Krylov subspace method. 
The residual vector can be defined 
both in the subspace diagonalization and shifted COCG method~\cite{TAKAYAMA2006} 
as 
\begin{eqnarray}
|r_j^{(\nu)} \rangle = (\varepsilon +{\rm i} \delta -{\hat H})|x_j^{(\nu)}\rangle -|j\rangle  ,
\label{residual}
\end{eqnarray}
where $|x_j^{(\nu)}\rangle$ is the $\nu$-th iterative solution. 
This residual vector can be monitored during the iterative calculation 
and we can stop the iterative procedure, 
without fixing the dimension of the Krylov subspace, 
once one can obtain the required accuracy. 
The norm of the residual vector can give the upper limit of the 
accuracy of the Green's function itself.~\cite{yamamoto2007b}
The shifted COCG method is numerically robust and 
one can reduce the norm of the residual vector to the machine accuracy. 
Therefore, the shifted-COCG method may be used 
to calculate accurate or fine density of electronic states 
in one-electron spectrum in  large-scale systems 
or the fine excitation spectra in many-electron problems. 
%%

%%%%%%%%%%%%%%%%%%%%%%%%%%%%%%%%%%%%%%%%%%%%%%%%%%%%%%%%%%%%%
%%%%%%%%%%%%%%%%%%%%%%%%%%%%%%%%%%%%%%%%%%%%%%%%%%%%%%%%%%%%%
\section{Applications}
%------------------------------------------------------------------------
%                              Fracture in Si crystals 
%------------------------------------------------------------------------
\subsection{Application to nano-scale systems}
%------------------------------------------------------------------------
\subsubsection{Formation and propagation of fracture 
in silicon crystal}\label{Si}\ \

In this subsection, we present an application study of our simulation; 
fracture formation and propagation phenomena 
in Si nano-scale crystal,~\cite{HOSHI2003,HOSHI2005}  
where the Hamiltonian is given as  a tight-binding representation.~\cite{KWON1994}  
The calculation was carried out by the Wannier-state methods 
with up to $10^5$ atoms. 
%%

%%%%%%%%%%%% figure 2 %%%%%%%%%%%%%%
\begin{figure}[htbp] 
\begin{center}
\includegraphics[width=0.8\linewidth]{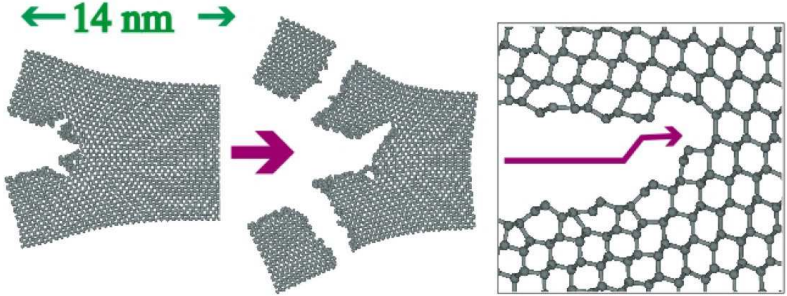}
\end{center}
\caption{\label{fig-Si} 
Silicon cleavage dynamics.~\cite{HOSHI2005}
A 14-nm-scale simulation result shows 
the bending of cleavage path from an unstable (001)-like plane 
into experimentally observed (111)-like and (110)-like planes. 
The right panel 
shows  a step-formation process on Si(111)-2x1 surface.
}
\end{figure}
%%%%%%%%%%%%%%%%%%%%%%%%%%%%%%%%%%%%%%%

%%
In the dynamical fracture formation process on the (001) plane, 
two bonds are broken and an asymmetric dimer ($2 \times 1$ periodicity on the 
resultant (001) surface) is formed after thermal motions of a time about 0.4~ps. 
First,  bonds are broken successively  
in an atom array of the dimer bonds on the plane along one of [110] directions. 
Along the formed asymmetric dimer bonds, the inter-atomic distance is shortened 
due to the formed bonding bonds. 
The distortion energy is accumulated  and, then,  
other bonds along a parallel atom array, but not the same, are broken. 
This fracture propagation 
(perpendicular to the direction of the formed asymmetric dimer bonds) 
is governed by the accumulated distortion energy. 
Our calculation can represent this surface breaking mechanism 
on the (001) plane of Si crystals.~\cite{HOSHI2003} 
We also studied with 14~nm scale Si crystals 
the easy-propagating plane of fracture.~\cite{HOSHI2005} 
It is widely known that the easy-propagating plane of fracture in Si 
is (110) or (111) planes. 
In case of fracture on the (111) plane, the (111)-$(2\times 1)$ surface reconstruction 
appears  (the Pandy structure~\cite{PANDEY}) and several steps are formed. 
The fracture propagation plane is not explained 
by the energy of established stable surfaces 
but by that of ideal or transient surface structure without reconstruction. 
In a MD process in a larger systems with 14nm length, 
even if a fracture propagation starts on a (001) plane, 
the plane of the fracture propagation changes to 
 (111) and (110) planes.  
Figure \ref{fig-Si} shows examples of the simulation results.
%%

%------------------------------------------------------------------------
%                              Liquid Carbon 
%------------------------------------------------------------------------
%%%%%%%%%%%% figure 3 %%%%%%%%%
\begin{figure}[htbp] 
\begin{center}
\includegraphics[width=0.5\linewidth]{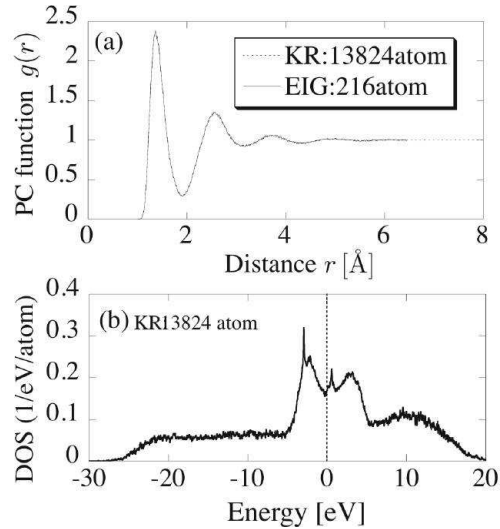}
\end{center}
\caption{\label{FIG-Liq-C} 
Molecular dynamical simulation of liquid carbon 
by the MD simulation with the Krylov subspace method. 
(a) The pair correlation function  and 
(b) the electron density of states.
}
\end{figure}
%%%%%%%%%%%%%%%%%%%%%%%%%%%%%%%%%%%%%%%

\subsubsection{Liquid carbon}\label{LiqC}\ \ 

Liquid carbon of 13,824 atoms was simulated with the Krylov subspace method.~\cite{HOSHI2006} 
The density and the temperature are set to
be $\rho=2.0~{\rm g~cm}^3$ and $T = 6000$~K. 
The time interval of a MD step  is  $\Delta t = 1$~fs 
and the subspace dimension and the number of interacting atoms are chosen to be 
$\nu = 30$ and $N_{\rm PR}= 200$, respectively. 
Figure \ref{FIG-Liq-C}(a) shows the resultant pair correlation (PC) function 
with comparison of the conventional eigenstate method of 216 atoms 
and we should notice that the two graphs are identical.
Figure~\ref{FIG-Liq-C}(b) shows the electronic density of states (DOS) 
of a system of with $13,824$ atoms 
from the Greenfs function, by the Krylov subspace method. 
The DOS calculation was achieved with the controlling parameters 
of a heavier computational cost 
($\nu = 300$ and $N_{\rm PR}= 1000$) and $\eta=0.05$~eV. 
Since the present Hamiltonian includes only s and p orbitals, 
the resultant DOS misses a structure in higher energy regions. 
The resultant DOS shows the characteristic profile of 
liquid carbon, 
e.g. a narrow $\pi$ band appears between $-5$ and $+5$~eV as in carbon nanotubes. 
The $\pi$ bond in the liquid phase is imperfect and
non-bonding (atomic) p states appear as a sharp peak near the chemical potential
($\varepsilon \simeq 0.6$~eV).
%%

%------------------------------------------------------------------------
%                              Gold nanowire 
%------------------------------------------------------------------------
%\subsection{Application to nano-scale systems \#2 : Helical multishell structure of gold nanowire}\label{Au}
\subsubsection{Helical multishell structure of gold nanowire}\label{Au} \ \

Another application is the formation process of helical multishell 
gold nanowires.~\cite{IGUCHI2007} 
Gold nanowires obtained by TEM thinning process 
have helical multishell structures 
along the original [110] axis with helicity, 
and the outermost shell is a (111)-like atomic sheet.~\cite{TAKAYANAGI2000} 
The difference of numbers of atoms between the outermost 
and  the next outermost shells is seven, called \lq magic number',
except cases of five and seven atoms on the outermost shell.
We proposed the two-stage formation  model of Au nanowires where  
the driving force for the helicity is the atom row slip. 
At the first stage, the outermost shell is dissociated 
from the inner shell to rotate freely. 
At the second stage, an atom row on the outermost shell slips  
and the (001) faces on the rod surface transform into (111) surfaces. 
%%

%%%%%%%%%%%%% figure 4 %%%%%%%%%%%%
\begin{figure}[htbp] 
\begin{center} 
\centerline{\hspace{3cm}
\resizebox{0.9\textwidth}{!}{
\includegraphics{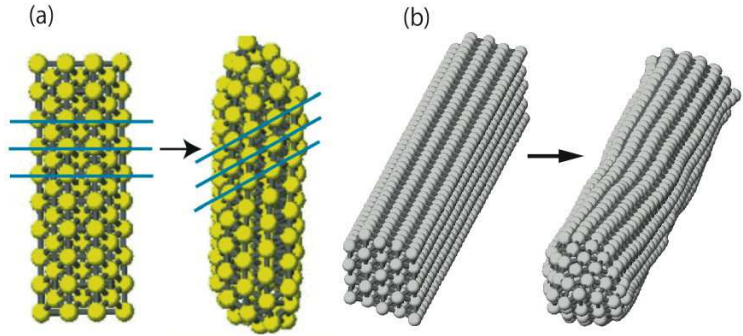}
}}
\end{center}
\caption{\label{FIG-Au} 
The formation process of the helical multishell structure in gold nanowires. 
Non-helical structure (left) is transformed into helical one (right).
(a)Formation process of (11-4) helical multishell gold nanowire.~\cite{IGUCHI2007}
The symbol (11-4) means that  
the number of atoms is eleven and four in the outer and inner shell, respectively. 
Atom row slip along the wire axis introduces the helicity as shown by three lines.  
(b) Transformation of structure of longer gold nanowire of 1,020 Au atoms 
and the resultant (15-8-1) nanowire contains several defects.
}
\end{figure}
%%%%%%%%%%%%%%%%%%%%%%%%%%%%%%%%%%%%%%%

%%
We verified the above two-stage model by using the MD simulations 
with a tight-binding  Hamiltonian,~\cite{KIRCHHOFF2001} 
starting from  an \lq ideal' nanowire of stacking (110) sections of the fcc lattice. 
The calculation was carried out by the eigen-state solver 
with about 80-1020 atoms. 
Here, we show results of 143 atoms in Fig.~\ref{FIG-Au}(a)~\cite{IGUCHI2007} 
and those in larger systems of 1020 atoms in (b). 
The total energy decreases almost monotonically after 1,000 MD steps 
(1MD step = 1fs).
First, the surface atoms dissociate from inner shell 
and, then, can move rather freely.  
From 2,000 to 5,000 MD steps, 
(001) sheet reconstructs into hexagonal (111)-like surface
with an atom-row slip deformation, and the helical structure on the surface appears. 
The inner shell rotates at the same time of the atom-row slip. 
Analysis of electronic structure shows that the mechanism in both 
stages is governed by the d-band electrons extending over the (111)-like surface, 
where the center of gravity of the d-band locates in the lower energy side.  
The helical nanowires appear only among metals with a wider d-band, 
e.g. in Au and Pt but not in Ag and Cu. 
Helicity is introduced by the surface reconstruction or the
atom-row slip on the (001) sheet, because the 
triangular (111)-like sheet is  more preferable  for d-orbitals
extending over the surface. 
The d-band width in platinum and gold is
commonly wider than that in lighter elements, Ag and Cu,
and the calculated result explains why platinum nanowire
can be also formed with helicity. 
%%

%------------------------------------------------------------------------
%                              multi-orbital extended Hubbard model 
%------------------------------------------------------------------------
\subsection{Application to many-electron systems : Excitation spectrum of 
multi-orbital extended Hubbard Hamiltonian on two-dimensional square lattice}\label{LSNO}

The transition metal oxides have been paid a great attention due to their 
various physical properties which are drastically changed 
and controllable by external fields or doping. 
Here we show an application of the shifted COCG method to the extended Hubbard model 
with doubly degenerated orbital and the inter-site Coulomb interaction 
on a two-dimensional square lattice.~\cite{yamamoto2007a} 
This is a model of La$_{\frac{3}{2}}$Sr$_{\frac{1}{2}}$NiO$_4$ 
and we used a finite unit of $N=8$ sites and the total number of electrons 
$N_{\rm e}=\frac{3}{2}N=12$.  
%%

%%%%%%%%%%%%% figure 5%%%%%%%%%%%
\begin{figure}[hpbt] 
\begin{center}
\resizebox{0.55\textwidth}{!}{
  \includegraphics{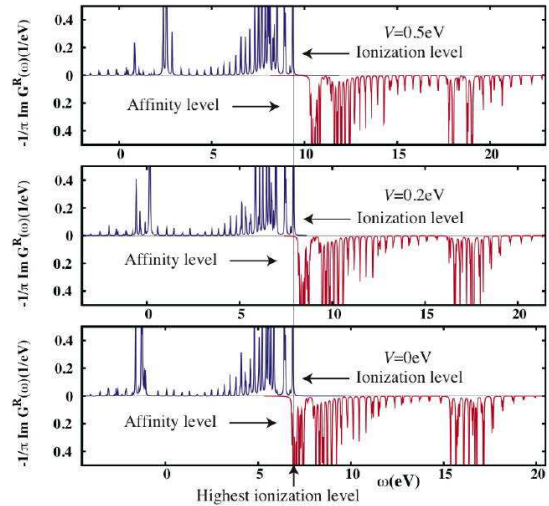}
}
\end{center}
\caption{\label{extHubbard} 
The excitation spectra, electron ionization and affinity, 
of doubly degenerated extended Hubbard model 
on $\sqrt{8}\times \sqrt{8}$ two-dimensional square lattice.~\cite{yamamoto2007a}
The matrix size of the reduced subspace of the total spin $S_z=0$ is $64,128,064$.
The excitation gap is formed with increasing the inter-site Coulomb interaction $V$.
}
\end{figure}
%%%%%%%%%%%%%%%%%%%%%%%%%%%%%%%%%%%%%%%

%%
We focused our attention to the Hilbert space of the total spin $S_z=0$ and 
the matrix size is $(_{16}C_6)^2=64,128,064$. 
The difficulty in many-electron systems is 
(1) the large dimension of the Hamiltonian matrix
which grows exponentially with increasing number
of atoms linearly and 
(2) very small energy intervals between adjacent eigenenergies 
which require a difficulty in separation of respective eigenvectors.
This difficulty requires fast, reliable and stable calculation algorithm 
for large matrices. 
Figure \ref{extHubbard} shows the excitation spectra of electron ionization and 
affinity levels and the energy gap between these two corresponds to the 
excitation gap. 
Normally, 
the Hubbard model of non-degenerate orbital, in case of the integer occupation, 
gives the insulating gap due to the on-site Coulomb interaction 
and, on the contrary, in case of non-integer occupation, the system is metal. 
Here, in the doubly degenerate case, the charge stripe order with an insulator gap 
is formed due to the inter-site Coulomb interaction $V$ and, on top of that, 
the spin stripe is formed with anisotropy of electron hoppings.~\cite{yamamoto2007a} 
Crucial point is that we should keep very high accuracy of the computation  
for judging the `gap', 
compared with the `level interval' in finite systems and 
that the iteration convergence should be controlled during the iterative calculation. 
Therefore, the capability of convergence (accuracy) monitoring and robustness 
are seriously important and 
the shifted COCG method can solve this difficulty. 
%%

%------------------------------------------------------------------------
%                              Conclusions 
%------------------------------------------------------------------------
\section{Conclusions}\label{Conclusion}

We have reviewed our recently developed methods 
for large-scale electronic structure calculation applied to both 
one-electron theory and many-electron theory. 
For large-scale systems of about 10's nm scale, 
one can use several solver methods simultaneously as a multi-solver method. 
We also explained differences between two theories from the viewpoint 
of large-scale matrix computation. 
Then we presented examples of the applications of nano-scale systems, 
the formation and propagation of fracture in large silicon crystals, 
the MD simulation in liquid carbon, and 
the formation of helical multishell structure of gold nanowires 
and an example of many-electron problems, 
the orbital degenerated extended Hubbard model. 
In these applications, we stressed the importance of 
the hybrid scheme of multiple solver methods and 
the novel computational algorithm. 
%%

%------------------------------------------------------------------------
%                              Acknowledgements 
%------------------------------------------------------------------------
\section*{Acknowledgments}

Numerical calculation was partly carried out 
using the supercomputer facilities of
the Institute for Solid State Physics, University of Tokyo. 

\section*{References}

%#########################################################################
%                               Bibliography
%#########################################################################
%-----------------------------------------------------------------


\begin{thebibliography}{}

\bibitem{CSE}
Comp. Sci. Eng., 
the January/February 2000 issue.  

\bibitem{SIAM} SIAM News, Vol.~33, No.~4.

\bibitem{KOHN96}
W. Kohn, Phys. Rev. Lett. {\bf 76}, 3168 (1996).

\bibitem{HOSHI2000}
T. Hoshi and T. Fujiwara, 
J. Phys. Soc. Jpn. {\bf 69}, 3773 (2000).

\bibitem{HOSHI2003}
T. Hoshi and T. Fujiwara, 
J. Phys. Soc. Jpn. {\bf 72}, 2429 (2003).

\bibitem{HOSHI2005}
T. Hoshi, Y. Iguchi and T. Fujiwara, 
Phys. Rev. B{\bf 72}, 075323 (2005).

\bibitem{TAKAYAMA2004}
R. Takayama, T. Hoshi and T. Fujiwara, 
J. Phys. Soc. Jpn. {\bf 73}, 1519 (2004). 


\bibitem{TAKAYAMA2006}
R. Takayama, T. Hoshi, T. Sogabe,  S-L. Zhang and T. Fujiwara, 
Phys. Rev. B{\bf 73}, 165108 (2006). 

\bibitem{geshi2004}
M. Geshi, T. Hoshi and T. Fujiwara, J. Phys. Soc. Jpn., {\bf 72}, 2880 (2004).

\bibitem{HOSHI2006}
T. Hoshi,  and T. Fujiwara, 
J. Phys: Condens. Matter. {\bf 18}, 10787 (2006). 

\bibitem{MAURI}
F. Mauri, G. Galli and R. Car, Phys. Rev. B{\bf 47}, 9973  (1993).

\bibitem{MARZARI}
N. Marzari and D. Vanderbilt, Phys. Rev. B{\bf 56}, 12847  (1997).

\bibitem{KWON1994}
I. Kwon, R. Biswas, C. Z. Wang, K. M. Ho and C. M. Soukoulis,   
Phys. Rev. B {\bf 49}, 7242 (1994).

\bibitem{XU}
C. H. Xu, C. Z. Wang, C. T.  Chan and K. M. Ho,  
J. Phys. Condens. Matter {\bf  4},  6047 (1992).

\bibitem{LMTO}
O. K. Andersen
and O. Jepsen, Phys. Rev. Lett. {\bf 53}, 2571 (1984). 

\bibitem{ELSES}
http://elses.jp

\bibitem{yamamoto2007a}
S. Yamamoto, T. Fujiwara and Y. Hatsugai, Phys. Rev. B{\bf 76}, 165114 (2007).

\bibitem{yamamoto2007b}
S. Yamamoto, T. Sogabe, T. Hoshi, S.-L. Zhang, and T. Fujiwara, 
in preparation.

\bibitem{Lanczos}
C. Lanczos, J. Res. Natl. Bur. Stand. {\bf 45}, 225 (1950);
{\it ibid.} {\bf 49}, 33 (1952).

\bibitem{CGmethod}
M. R. Hestenes and E. Stiefel, J. Res. Natl. Bur. Stand. {\bf 49}, 409 (1952).

\bibitem{FROMMER2003}
A. Frommer, Computing {\bf 70}, 87 (2003).  

\bibitem{SOGABE2007}
T. Sogabe, T. Hoshi, S.-L. Zhang, and T. Fujiwara, 
Frontiers of Computational Science, pp. 189-195, 
Ed. Y. Kaneda, H. Kawamura and M. Sasai, 
Springer Verlag, Berlin Heidelberg
(2007). 

\bibitem{PANDEY}
K. C. Pandey, Phys. Rev. Lett. {\bf 47}, 1913 (1981).

\bibitem{IGUCHI2007}
Y. Iguchi, T. Hoshi and T. Fujiwara, 
Phys. Rev. Lett. {\bf 99}, 125507 (2007). 

\bibitem{TAKAYANAGI2000}
Y. Kondo and K. Takayanagi, Science {\bf 289}, 606 (2000).

\bibitem{KIRCHHOFF2001}
F. Kirchhoff, M. J. Mehl, N. I. Papanicolaus, D. A. Papaconstantpoulos and 
F. S. Khan, 
Phys. Rev. B{\bf 63}, 195101 (2001).



%-----------------------------------------------------------------
\end{thebibliography}
\end{document}